\documentstyle[12pt,amssymb]{article}
\begin{document}
\begin{titlepage}

\begin{center}
{\Large\bf Holomorphic Quantization on the Torus and Finite Quantum Mechanics}

\vskip1.5truecm
G. G. Athanasiu$^{*}$\footnote{e-mail: athanasi@iesl.forth.gr.}, 
E. G. Floratos$^{**}$\footnote{e-mail: floratos@cyclades.nrcps.ariadne-t.gr.
On leave of absence from Physics Department, University of Crete.}
 and S. Nicolis$^{***}$\footnote{e-mail: nicolis@celfi.phys.univ-tours.fr}

\vskip1truecm 

{\sl CNRS--Laboratoire de Physique Th\'eorique 
de l'Ecole Normale Sup\'erieure\footnote{Unit\'e propre du CNRS (UPR 701) associ\'ee \`a l'ENS et \`a l'Universit\'e Paris-Sud.}\\
24 rue Lhomond, 75231 Paris Cedex 05, France}

\vskip1truecm

$^{*}${\sl Physics Department, University of Crete\\
{\rm and}\\
F.O.R.T.H., Heraklion, Crete, Greece}

\vskip1truecm 

$^{**}${\sl I. N. P. , NRCPS ``Demokritos''\\
15310 Aghia Paraskevi, Athens, Greece} 

\vskip1truecm

$^{***}${\sl CNRS--Laboratoire de Math\'ematiques et de Physique Th\'eorique (EP93)\\
D\'epartement de Physique, Universit\'e de Tours\\
Parc Grandmont, 37200 Tours, France}

\end{center}

\vskip1truecm

\begin{abstract}
{\small 
We construct explicitly the quantization of classical linear maps of 
$SL(2,{\Bbb R})$ on toroidal phase space, of arbitrary modulus, using the 
holomorphic (chiral) version of the metaplectic representation. We show that 
Finite Quantum Mechanics (FQM) on tori of arbitrary integer discretization, 
is a consistent restriction of the holomorphic quantization of 
$SL(2,{\Bbb Z})$ to the subgroup $SL(2,{\Bbb Z})/\Gamma_l$, $\Gamma_l$ being 
the principal congruent subgroup mod $l$, on 
a finite dimensional Hilbert space. The generators of the ``rotation group'' mod $l$, 
$O_{l}(2)\subset SL(2,l)$, for arbitrary values of $l$ are determined as well as their quantum mechanical 
eigenvalues and eigenstates. }
\end{abstract}

\end{titlepage}
\section{Introduction}
A most fascinating branch of mathematics, Number Theory~\cite{numbertheory}, quite unexpectedly, has made its 
appearance in a variety of research areas in physics the last fifteeen years. In classical and 
quantum chaos~\cite{qc2,qc3,qc4,qc5}, localization in incommensurate lattices~\cite{loc}, classification of rational 
conformal field theories~\cite{rct1,rct2}, in string theory~\cite{st},
 etc.~(cf. also ref.~\cite{leshouches}). 

On the other hand, Number Theory, as is well known, is, for many years, an important tool in theoretical computer 
science (algorithms, cryptography) as well as in signal processing~\cite{sigproc1,sigproc2}. 

Motivated by recent work, on the information paradox of black holes~\cite{bh}, as well as by indications that 
string theory predicts an absolute minimum distance in nature, of the order of 
$M_{\rm P\rm l\rm a\rm n\rm c\rm k}^{-1}\approx 10^{-33}\,{\rm c\rm m}$~\cite{21-24}, 
two of the authors 
reconsidered the ancient question, why nature has to use 
real and complex numbers, once there is a fundamental unit of length, and they proposed to study quantum mechanics over 
finite sets of integers with the structure of finite algebraic fields, 
hoping to be able, eventually, to formulate 
field and string theory over them~\cite{afetal}, which possess the property 
of containing finite information per unit of physical volume.  
The basic obstacle here is that these number fields cannot accommodate 
metric structures and the notion of dimension. The very difficult task, then, is to reproduce, at scales much larger than 
the Planck scale, quantum physics as we know it. 

In this note we take a first step in connecting Finite Quantum Mechanics
FQM~\cite{bi} to a continuum quantum mechanics of a rather particular type. 
Indeed we show that FQM is a consistent and {\em exact} discretization of holomorphic quantum mechanics on toroidal 
phase spaces for arbitrary moduli, thereby establishing a possible link to rational conformal field theories on the 
torus. We extend the work of ref.~\cite{afetal} to torus discretizations of any length. 

The plan of the paper is as follows: in the next section we review holomorphic 
quantum mechanics on the (continuum) torus; we then discuss finite quantum 
mechanics and end by discussing some properties of harmonic oscillator 
eigenfunctions on these spaces and further perspectives.

\section{Holomorphic Quantum Mechanics}
 
We start by describing holomorphic quantum mechanics on the torus~\cite{cartier} (cf. also Leboeuf and Voros in ref.~\cite{qc5}). The torus of complex modulus $\tau\in{\Bbb C}$
is defined as the coset space $\Gamma={\Bbb C}/{\Bbb L}$, where ${\Bbb L}$ is the integer lattice 
${\Bbb L}=\{m_1+\tau m_2| (m_1,m_2)\in{\Bbb Z}\times{\Bbb Z}\}$. The torus $\Gamma$ is the set of points of the complex 
plane ${\Bbb C}, \,z=q+\tau p, \,\,q,p,\in [0,1]$. The symplectic structure of  ${\Bbb C}$ induces on $\Gamma$ the 
(symplectic) form
\begin{equation}
\label{sympl}
\Omega=-\frac{1}{2{\rm i}}dz\wedge d\overline{z}=\tau_2dq\wedge dp,
\end{equation}
where $\tau=\tau_{1}+{\rm i}\tau_2$. The corresponding group of symplectic transformations is $SL(2,{\Bbb R})$ acting on 
$(q,p)\,{\rm mod}\,1$~\cite{qc2}. To define holomorphic quantum mechanics on $\Gamma$ we start by the classical evolution 
in the phase space $\Gamma$ 
under elements of $SL(2,{\Bbb R})$. The most general quadratic Hamiltonian 
\begin{equation}
\label{Ham}
{\cal H}=\frac{\tau_2}{2}(q\, ,\, p)\left(\begin{array}{cc}
                                       -c & a\\
                                        a & b \\
                                     \end{array}\right)\left(\begin{array}{c}
                                                                   q \\
                                                                   p\\
                                                               \end{array}\right)
\end{equation}
leads to the evolution equations 
\begin{equation}
\label{evol}
\frac{d}{dt}(q\,\, ,\,\, p)=(q\,\, ,\,\, p)\left(\begin{array}{cc}a & c \\ b & -a \\ \end{array}\right)
\end{equation}
which are immediately integrated to
\begin{equation}
\label{evolint}
(q(t)\,\, ,\,\, p(t))\equiv(q(0)\,\, ,\,\, p(0)){\cal R}(t)\,{\rm mod}\,1
\end{equation}
with ${\cal R}(t)\in SL(2,{\Bbb R})$ given by 
\begin{equation}
\label{evolR}
{\cal R}(t)={\exp}\left[{\displaystyle t}
\left(\begin{array}{cc} a & c \\ b & -a \\ \end{array}\right)\right]
\end{equation}
The {\em quantum mechanical} evolution, with Weyl ordering as in eq.~(\ref{Ham}), is also simple and leads to 
\begin{equation}
\label{evolRq}
(\hat{q}(t)\,\, ,\,\, \hat{p}(t))=(\hat{q}(0)\,\, ,\,\,\hat{p}(0)){\cal R}(t)
\end{equation}
The position and momentum operators, $\hat{q}$ and $\hat{p}$ satisfy 
\begin{equation}
\label{commute}
[\hat{q},\hat{p}]=\frac{{\rm i}\hbar}{\tau_2}
\end{equation}

From (\ref{evolRq}) and the Heisenberg equations of motion we have that 
\begin{equation}
\label{U}
{\cal U}(t)(\hat{q}(0)\,\, ,\,\,\hat{p}(0)){\cal U}^{-1}(t)=
(\hat{q}(0)\,\, ,\,\,\hat{p}(0)){\cal R}(t)
\end{equation}
with ${\cal U}(t)$ the evolution operator
\begin{equation}
\label{Uevol}
{\cal U}(t)=\exp\left[
{\displaystyle\frac{{\rm i}t}{\hbar}
 \frac{\tau_2}{2}(\hat{q}\,\, ,\,\,\hat{p})}
\left(\begin{array}{cc}-c & a\\a & b\\ \end{array}\right)
\left(\begin{array}{c}\hat{q}\\ \hat{p}\\ \end{array}\right)\right]
\end{equation}     
This last relation, as we shall see, defines a representation of $SL(2,{\Bbb R})$. 
The Hilbert space of quantum mechanics on the torus $\Gamma$ consists of functions (to be more precise this is a space of sections of a $U(1)$ 
bundle over $\Gamma$) given by series 
\begin{equation}
\label{functions}
f(z)=\sum_{n\in{\Bbb Z}} c_{n}{\rm e}^{{\rm i}\pi n^2\tau + 2\pi{\rm i}nz}
\end{equation}
with norm~\cite{mumford}
\begin{equation}
\label{norm}
||f||^2=\int {\rm e}^{-2\pi y^2/\tau_2} |f(z)|^2 dxdy,\,\,\,\tau_2>0
\end{equation}
On this space consider the action of the operators ${\cal S}_{b}$ and ${\cal T}_{a}$ 
\begin{eqnarray}
\label{ST}
({\cal S}_{b}f)(z) & = & f(z+b),\,\,\,\forall f\in {\Bbb H}_{\Gamma}\nonumber\\
({\cal T}_{a}f)(z) & = & {\rm e}^{{\rm i}\pi a^2\tau+2\pi{\rm i}z}f(z+a\tau),\,\,\,a,b\in{\Bbb R}
\end{eqnarray}
which satisfy the fundamental Weyl commutation relations (CR), the integrated 
form of Heisenberg CR,
\begin{equation}
\label{HCR}
{\cal S}_{b}{\cal T}_{a}={\rm e}^{2\pi{\rm i}ab}{\cal T}_{a}{\cal S}_{b} 
\end{equation}
The operators ${\cal S}$ and ${\cal T}$ are so chosen that the classical 
Jacobi theta function~\cite{mumford} 
\begin{equation}
\label{Jacobi}
\theta(z|\tau)=\sum_{n\in{\Bbb Z}}{\rm e}^{{\rm i}\pi n^2\tau+2\pi{\rm i}nz}
\end{equation}
is invariant under ${\cal S}_{1}$ and ${\cal T}_{1}$.

The space ${\Bbb H}_{\Gamma}$ carries an infinite dimensional, unitary, irreducible representation of the Heisenberg group defined as
\begin{equation}
\label{Heisenberg}
{\cal W}(\lambda,a,b)f=\lambda{\cal T}_{a}{\cal S}_{b}f,\,\,\,\lambda\in U(1),a,b\in{\Bbb R},\,\,\forall f\in {\Bbb H}_{\Gamma}
\end{equation}
with composition law
\begin{equation}
\label{compHeis}
{\cal W}(\lambda,a,b){\cal W}(\lambda^{'},a^{'},b^{'})=
{\cal W}(\lambda\lambda^{'}{\rm e}^{2\pi{\rm i}ba^{'}},a+a^{'},b+b^{'})
\end{equation}
In holomorphic quantum mechanics on the torus~\cite{cartier}, $\hat{q}$ and $\hat{p}$ are given by
\begin{equation}
\label{holoqm}
\hat{q}=-{\rm i}\partial_{z}, \,\,\,\hat{p}=-2\pi z+{\rm i}\tau\partial_{z}
\end{equation}
and thus
\begin{eqnarray}
\label{holost}
{\cal S}_{1}&=& {\rm e}^{{\rm i}\hat{q}}\nonumber\\
            & &                         \nonumber\\             
{\cal T}_{1}&=& {\rm e}^{-{\rm i}\hat{p}}
\end{eqnarray}
where we have chosen $\hbar=2\pi\tau_2$.  

We are ready now to describe the 
metaplectic representation of $SL(2,{\Bbb R})$ on 
the space ${\Bbb H}_{\Gamma}$. 
For every $(q,\,p)\in\Gamma$ the evolution operator, ${\cal U}(t)$,
(cf. eq.~(\ref{Uevol})), satisfies the relation (cf. eqs.~(\ref{U}),(\ref{evolR}))
\begin{equation}
\label{holoU}
{\cal U}_{{\cal R}}^{-1}(t){\cal J}_{q,p}{\cal U}_{\cal R}(t)={\cal J}_{(q\,\,p){\cal R}(t)}
\end{equation}
where 
\begin{equation}
\label{heiselem}
{\cal J}_{q,p}\equiv {\rm e}^{\displaystyle {\rm i}(-q\hat{p}+p\hat{q})}
\end{equation}
 is an element of the Heisenberg group acting on ${\Bbb H}_{\Gamma}$. 

The metaplectic representation~\cite{cartier,weil} of $SL(2,{\Bbb R})$
is defined by eq.~(\ref{holoU}) and, in general, is a 
projective representation.

\section{Finite Quantum Mechanics}
We now recall the basic features of  FQM and its relation to the 
holomorphic QM. 

The torus phase space has been the simplest prototype for studying classical
and quantum chaos~\cite{qc2,qc3,qc4,qc5}. Discrete 
elements of $SL(2,{\Bbb R})$, i.e. elements of the modular group 
$SL(2,{\Bbb Z})$, are studied on discretizations of the torus with rational 
coordinates of the same denominator $l$, $(q,p)=(n_1/l,n_2/l)\in\Gamma,\,
n_1,n_2,l\in{\Bbb Z}$ and their periodic trajectories mod 1 are examined studying 
the periods of elements ${\cal A}\in SL(2,{\Bbb Z})$ mod $l$. The action mod 1 
becomes mod $l$ on an equivalent torus, $(n_1,n_2)\in l\Gamma$. The classical motion of such discrete dynamical systems is usually ``maximally''
disconnected and chaotic~\cite{qc3,qc5}.

FQM is the  quantization of these discrete linear maps and the corresponding 
one-time-step
evolution operators ${\cal U}_{\cal A}$ are $l\times l$ unitary matrices called 
{\em quantum maps}. In the literature~\cite{qc4,qc5}
these maps are determined semi-classically. In ref.~\cite{afetal,bi} 
the exact quantization of $SL(2,{\Bbb F}_p)$, 
where ${\Bbb F}_{p}$ is the simplest finite field of $p$ elements with 
$p$ a prime number was studied in detail.
 In the following we shall extend the results of 
ref.~\cite{afetal} to $l=p^n$ and we shall discuss the case of 
arbitrary integer $l$.

Consider the  subspace ${\Bbb H}_{l}(\Gamma)$ of ${\Bbb H}_{\Gamma}$ 
with periodic Fourier coefficients $\{c_n\}_{n\in{\Bbb Z}}$ of period $l$
\begin{equation}
\label{fourier}
c_n=c_{n+l}\,\,\,n\in {\Bbb Z},\,\,l\in{\Bbb N}
\end{equation}
The space   ${\Bbb H}_{l}(\Gamma)$ is $l$-dimensional and there is a discrete
Heisenberg group~\cite{Weyl}, 
with generators ${\cal S}_{1/l}$ and ${\cal T}_{1}$ 
acting as~\cite{cartier,mumford}
\begin{eqnarray}
\label{STl}
({\cal S}_{1/l}f)(z) &=& \sum_{n\in{\Bbb Z}}c_n
{\rm e}^{2\pi{\rm i}n/l}{\rm e}^{2\pi{\rm i}nz+\pi{\rm i}n^2\tau}\nonumber\\
 & & \nonumber\\
({\cal T}_{1}f)(z) &=& \sum_{n\in{\Bbb Z}}c_{n-1}
{\rm e}^{2\pi{\rm i}nz+\pi{\rm i}n^2\tau},\,\,\,c_n\in{\Bbb C}
\end{eqnarray}
On the $l$-dimensional subspace of vectors $(c_1,\ldots,c_l)$ the two 
generators are represented by
\begin{eqnarray}
\label{gener}
({\cal S}_{1/l})_{n_1,n_2}=Q_{n_1,n_2}=\omega^{(n_1-1)}\delta_{n_1,n_2}
\nonumber\\
\nonumber   \\
({\cal T}_{1})_{n_1,n_2}=P_{n_1,n_2}=\delta_{n_1-1,n_2}
\end{eqnarray}
with $\omega=\exp(2\pi{\rm i}/l)$. The Weyl relation becomes 
\begin{equation}
\label{Weyl}
QP=\omega PQ
\end{equation}
and the Heisenberg group elements are
\begin{equation}
\label{heisel}
{\cal J}_{r,s}=\omega^{r\cdot s/2}P^rQ^s
\end{equation}
In the literature 
the metaplectic representation of $SL(2,l)$, (the group of $2\times 2$,
integer valued 
matrices mod $l$), is known for 
$l=p^n$~\cite{tanaka}\footnote{The representation theory of the symplectic 
group $SL(2,{\Bbb F}_{p^n})$ may be found in ref.~\cite{tanaka2}.}

The Weyl-Fourier form of ${\cal U}_{\cal A}$ is~\cite{bi}
\begin{equation}
\label{weylfour}
{\cal U}_{\cal A}=\frac{\sigma(1)\sigma(\delta)}{p}
\sum_{r,s=0}^{p-1} {\rm e}^{\displaystyle\frac{2\pi{\rm i}}{p}
[br^2+(d-a)rs-cs^2]/2\delta}
{\cal J}_{r,s}
\end{equation}
where 
\begin{eqnarray}
\label{sl2defs}
{\cal A} &=&\left(\begin{array}{cc} a & b\\c &d\end{array}\right)\in SL(2,{\Bbb F}_{p}),\,\,\,\delta=2-a-d\nonumber\\
 & & \nonumber\\
\sigma(a)&=&\frac{1}{\sqrt{p}}\sum_{r=0}^{p-1}\omega^{ar^2}=
(a|p){\frak p}
\end{eqnarray}
$(a|p)$ is the Jacobi symbol~\cite{numbertheory} and
$$
{\frak p}=\left\{\begin{array}{cc}1 & p=4k+1\\{\mathrm i} & p=4k-1
\\ \end{array}\right.
$$

All the operations in the exponent are carried out in the field ${\Bbb F}_{p}$.
If $\delta\equiv 0$ mod $p$ we use the trick
\begin{equation}
\label{trick}
\left(\begin{array}{cc}a & b\\c & d\\ \end{array}\right) = 
\left(\begin{array}{cc} 0 & 1 \\ -1 & 0\\ \end{array}\right)
\left(\begin{array}{cc} -c & -d \\ a & b\\ \end{array}\right)
\end{equation}
and the fact that ${\cal U}_{\cal A}$ is a representation (cf. ref.~\cite{bi}). 

In ref.~\cite{afetal} the eigenproblem for the generators of the 
``rotation subgroup'' of $SL(2,{\Bbb F}_{p})$, $O_{2}(p)$, was solved and 
an explicit list of the generators ${\cal R}_{0}$ for primes $p<20\, 000$ was given.
The spectrum of   ${\cal R}_{0}$ is linear and all the eigenvectors, which 
are real, were found analytically for primes $p=4k+1$. 
In fact they are appropriately
weighted Hermite polynomials over the finite field ${\Bbb F}_{p}$~\cite{afetal,evans}. 
All of them 
 turned out to be extended, in the sense that their support is the 
full set ${\Bbb F}_{p}$ and their components randomly distributed.

The first step to generalize the results of ref.~\cite{afetal} for $O_{l}(2)$
 is to consider 
integers $l=p^n$, powers of primes. We shall need the explicit form of 
${\cal U}_{\cal A}$ for $l=p$ because this is immediately generalized to 
$l=p^n$ 
\begin{equation}
\label{UA}
({\cal U}_{\cal A})_{n_1,n_2}=\frac{1}{\sqrt{p}}(-2c|p)
{\frak p}
\omega^{-[a(n_1-1)^2+d(n_2-1)^2-2(n_1-1)(n_2-1)]/2c}
\end{equation}
for $c\not\equiv 0$ mod $p$ (otherwise apply eq.~(\ref{trick})).

Imposing eq.~(\ref{weylfour}) for $l=p^n$ we need the 
Gau\ss~sum\footnote{A complete study of this sum for arbitrary integer $l$ can be found in the chapter 
``Cyclotomic Fields'' of S. Lang in ref.~\cite{numbertheory}; cf. also ref.~\cite{auslander}.} 
\begin{equation}
\label{gausssum}
{\frak G}(k,l)=\frac{1}{\sqrt{l}}\sum_{r=0}^{l-1}
{\rm e}^{2\pi{\rm i}kr^2/l}
\end{equation}
It enjoys the property
\begin{equation}
\label{gaussprop}
{\frak G}(k,p^n)=p{\frak G}(k,p^{n-2})
\end{equation}
which implies that
\begin{equation}
\label{evenn}
{\frak G}(k,p^{2m})=p^m
\end{equation}
so, for $n=2m$,
\begin{equation}
\label{matelemU}
({\cal U}_{\cal A})_{n_1,n_2}=\frac{1}{\sqrt{p^{2m}}}
\exp\left[-\left(\frac{2\pi{\rm i}}{p^{2m}}
\left[a(n_1-1)^2+d(n_2-1)^2-2(n_1-1)(n_2-1)\right]/2c\right)\right]
\end{equation}
with $c\not\equiv$ 0  mod $p$. For odd powers of $p$, $n=2m+1$, 
\begin{equation}
\label{oddn}
{\frak G}(k,p^{2m+1})=p^m{\frak G}(k,p)
\end{equation}
so we have only to replace $p$ by $p^{2m+1}$ in eq.~(\ref{UA}) and 
$1/2c$ is taken mod $p^{2m+1}$. 

The above results can be deduced also from the work of S. Tanaka on the 
representations of $SL(2,p^n)$~\cite{tanaka}. 

For practical calculations of spectra and eigenvectors of $O_{p^n}(2)$
for various primes one has to determine the corresponding  generators 
${\cal R}_{0}$. Here we explicitly present their construction. 
In ref.~\cite{afetal} ${\cal R}_{0}$ was found in the case of $p=4k+1$
once a primitive element of ${\Bbb F}_{p}$, ${\frak g}$, is given.
\begin{equation}
\label{R0}
{\cal R}_{0}=\left(\begin{array}{cc}\displaystyle
\frac{{\frak g}+{\frak g}^{-1}}{2} & 
\displaystyle \frac{{\frak g}^{-1}-{\frak g}}{2\frak t}\\
\displaystyle \frac{{\frak g}-{\frak g}^{-1}}{2\frak t} &
\displaystyle \frac{{\frak g}+{\frak g}^{-1}}{2}\\
\end{array}\right)
\end{equation}
here ${\frak t}\equiv {\frak g}^k$ mod $p$, ${\frak t}^2\equiv -1$ mod $p$ 
and all operations in the entries of eq.~(\ref{R0}) are performed mod $p$. 

The set of integers mod $p^n$ does not form a finite field, but there is 
a multiplicative subgroup, composed of all the integers,
 $\not\equiv 0$~mod~$p$. 
A known theorem states that, if ${\frak g}^{p-1}\not\equiv 1$ mod $p^n$, then
${\frak g}$ is a generator of this cyclic group with order $\phi(p^n)=p^n-p^{n-1}$. If $p=4k+1$, $\phi(p^n)$ is divisible by 4 and there is an element 
${\frak t}$ (${\frak t}^2\equiv -1$ mod $p^n$), ${\frak t}\equiv 
{\frak g}^{\phi(p^n)/4}$. In this case ${\cal R}_{0}$ is given by eq.~(\ref{R0}) where all the operations are mod $p^n$. 

In the case $p=4k-1$ we need to know a primitive element 
${\sf w}={\sf w}_1+{\rm i}{\sf w}_2$ of 
${\Bbb F}_{p^2}$~\cite{bi,afetal}. The corresponding generator of $O_{p}(2)$
is
\begin{eqnarray}
\label{R0-1}
{\cal R}_{0}=\left(\begin{array}{cc}u_1&u_2\\-u_2&u_1\\ \end{array}\right)
\nonumber\\
u_1+{\rm i}u_2=\frac{{\sf w}^2}{\frak g},\,\,\,{\frak g}={\sf w}\overline{\sf w}\in{\Bbb F}_{p}
\end{eqnarray} 
here $\overline{\sf w}={\sf w}_1-{\rm i}{\sf w}_2\equiv {\sf w}^p$ mod $p$
and ${\frak g}$ can be shown to be a primitive element of ${\Bbb F}_{p}$. 
A list of ${\cal R}_{0}$ and ${\sf w}$ for all primes $p=4k-1<20\,000$ 
can be found in ref.~\cite{afetal}. 

For $l=p^n$, $p=4k-1$, we can find primitive elements ${\sf w}\in {\Bbb F}_{p^2}$ such that ${\frak g}={\sf w}\overline{\sf w}$ has the property 
${\frak g}^{p-1}\not\equiv 1$ mod $p$ and the 
corresponding generator ${\cal R}_{0}$ is given by eq.~(\ref{R0-1}) where 
all the operations are performed mod $p^n$. 

From the above one can find that, for $l=p^n=(4k+1)^n$ the period of 
the generator is $\phi(p^n)=p^n-p^{n-1}$, while, for $l=p^n=(4k-1)^n$, 
the period is $p^n+p^{n-1}$. 

For arbitrary $l=\prod_{i=1}^{s} p_{i}^{n_i}$ 
$SL(2,l)=\otimes_{i=1}^{s}SL(2,p_{i}^{n_i})$~\cite{numbertheory,rct1},
It is known that $SL(2,l)$ is the coset space $SL(2,{\Bbb Z})/\Gamma_{l}$, 
where ${\Gamma}_l$ is the set of matrices ${\cal A}\in SL(2,\Bbb Z)$, such that 
${\cal A}=\pm I$ mod $l$. This is a normal subgroup of $SL(2,\Bbb Z)$ and is called 
the {\em principal congruent subgroup mod l}. It plays an important role 
in the geometry of Riemann surfaces and the classification of modular 
forms~(cf. the article by D. Zagier in ref.~\cite{leshouches}).   
$SL(2,l)$ consists of {\em nested} sequences, in the sense that 
$SL(2,l)\subset SL(2,l^{'})$ when $l^{'}\equiv 0$ mod $l$.

The metaplectic representation, eq.~(\ref{weylfour}), can be extended to any $l$, once $\delta=2-a-d\not\equiv 0$ mod $p_i$ (for any $p_i$); the Gau\ss\/ sums 
can be easily evaluated (cf. S. Lang in ref.~\cite{numbertheory})--unfortunately, there isn't any simple, {\em unique} answer for arbitrary ${\cal A}\in SL(2,l)$. 
For the class of ${\cal A}$'s, of the form
\begin{eqnarray}
\label{classA} 
{\cal A}=\left(\begin{array}{cc}even&odd\\odd&even\\ \end{array}\right)
 &{\mathrm o\mathrm r}&
{\cal A}
=\left(\begin{array}{cc}odd&even\\even&odd\\ \end{array}\right)
\end{eqnarray}
Hannay and Berry~\cite{qc4} have written down the semiclassical form of 
${\cal U}_{\cal A}$. It is not difficult to see that the 
metaplectic representation, eq.~(\ref{weylfour}),leads to the same results. 
For the other forms of ${\cal A}$ the answer for ${\cal U}_{\cal A}$ does not have the 
same form for all $l$. Our main interest is the harmonic oscillator subgroup,
$O_{l}(2)\subset SL(2,l)$. As we mentioned before, $SL(2,l)$ can be decomposed
into a tensor product of $SL(2,p_i^{n_i})$, $i=1,\ldots,s$ 
over the prime factors of $l$. 
The same happens for $O_{l}(2)$, which is an abelian group, with $s$ 
cycles and with generators ${\cal R}_{0}(p_i^{n_i})$. Its representations 
may thus be obtained by tensoring powers of ${\cal U}_{{\cal R}_{0}(p_i^{n_i})}$. 

\section{Perspectives}
We discuss finally the construction of the eigenstates of the harmonic oscillator subgroup (mod $l$). These are presumably the building blocks of field theories (and string theories) on discretized toroidal phase spaces. 
It is enough to determine the eigenstates (and eigenvalues) of $O_{p^{n}}(2)$
for any prime $p$ and (positive) integer $n$.
 From the construction of ${\cal R}_{0}$ and their diagonalized form
\begin{equation}
\label{diagR}
\left(\begin{array}{cc}a &b\\-b&a\\ \end{array}\right)={\cal L}
\left(\begin{array}{cc}a-{\frak t}b & 0\\ 0&a+{\frak t}b\\ \end{array}\right)
{\cal L}^{-1}\,\,\,\,\,\,a^2+b^2\equiv\,1\,{\mathrm m\mathrm o\mathrm d}\,p^n
\end{equation}
where 
\begin{equation}
\label{diagL}
{\cal L}=\frac{1}{2\frak t}\left(\begin{array}{cc}1 & 1\\ -\frak t & \frak t
\end{array}\right)
\end{equation}
with ${\frak t}^2\equiv -1\,{\rm mod}\,p^n$, 
$\sqrt{2\frak t}\equiv (1+{\frak t})\,{\rm mod}\,p^n$, we diagonalize the corresponding 
${\cal U}_{\Delta}$, where
$$
\Delta=\left(\begin{array}{cc}a-{\frak t}b & 0\\ 0&a+{\frak t}b\\ \end{array}\right)
$$
As was shown in ref.~\cite{afetal}, ${\cal U}_{\Delta}$ is a 
{\em circulant} matrix
 for $l=p$
(and the same happens here as well, in each sector 
$p^n$) for $p=4k+1$: 
its first row is $e_1=(1,0,\ldots,0)$ and each subsequent row 
has the element $1$ shifted by $\frak g^{-1}$ mod $p^n$ positions to the right from the last. The eigenvectors of ${\cal U}_{\Delta}$ are easily 
found to be the multiplicative characters of the set of integers 
mod $p^n$, extending the results of ref.~\cite{afetal}.
The eigenvalues of ${\cal U}_{\Delta}$ are roots of unity, of order $p^n-p^{n-1}$ and
$p^{n-1}$ of them must be degenerate.

For $p=4k-1$, it is possible to find {\em directly} the corresponding eigenvectors 
of ${\cal U}_{\Delta}$. They are the multiplicative characters of the 
rotation group $O_{p^n}(2)$, while the eigenvalues 
are roots of 
unity of order $p^n+p^{n-1}$.

We close this note by writing ${\cal U}_{\cal A}$, ${\cal A}\in SL(2,l)$, in terms of 
holomorphic operators on ${\Bbb H}_{l}(\Gamma)$. Define the elements 
of the Heisenberg group
\begin{equation}
{\cal J}_{r,s}=\exp\left({\rm i}\left[-r\hat{p}+s\hat{q}\right]\right),\,\,
r,s=1,\ldots,l
\end{equation}
with 
\begin{eqnarray}
\hat{p}&=&-2\pi z+{\rm i}\tau\partial_z\nonumber\\
\hat{q}&=&\frac{-{\mathrm i}}{l}\partial_z\nonumber\\
\left[\hat{q},\hat{p}\right] &=&\frac{2\pi{\mathrm i}}{l}
\end{eqnarray}
In eq.~(\ref{weylfour}) we substitute $P$ and $Q$ with ${\cal T}_{1}$ and
${\cal S}_{1/l}$ respectively and carry out first the summation over $s$ and 
then over $r$. We end up with
\begin{equation}
\label{UAfinal}
{\cal U}_{\cal A}=\exp\left(-\frac{2\pi{\mathrm i}}{l}\frac{\delta}{2b}
\left(\frac{l}{2\pi}\hat{p}\right)^2\right)
\exp\left(-\frac{2\pi{\mathrm i}}{l}\frac{1}{2b}\left[(1-a)\frac{l}{2\pi}\hat{p}+b\frac{l}{2\pi}\hat{q}\right]^2\right)
\end{equation}
where the operators in the exponents have integer eigenvalues. We assume 
$\delta,b\not\equiv 0$ mod $p$.   

Finally we address the issue of localization of the eigenstates of 
${\cal U}_{{\cal R}_0}$ for the harmonic oscillator. For $l=p^n$ this operator is represented
by a $p^n\times p^n$ unitary matrix of period $p^n\mp p^{n-1}$ for $p=4k\pm 1$.


Higher powers of ${\cal R}_0$ (higher degeneracy but {\em smaller} period) have classical orbits that are
localized in phase space (intuitively understandable: since the period is smaller the orbits wander less in phase space)--and the quantum eigenstates follow suit. A nice 
example is provided by the finite Fourier transform; set ${\frak F}={\cal R}_{0}^{\phi(p^n)/4}$, with ${\frak F}^4=I$. The quantum map ${\cal U}_{\frak F}$, the 
finite Fourier transform, is known to possess localized eigenstates~\cite{mehta}
\begin{equation}
\label{fifour}
{\varphi}_k(j)=\left.\left(\frac{\partial}{\partial x}\right)^k 
\left[e^x
\theta\left(\left.\frac{j}{l}-x\sqrt{\frac{2}{\pi l}}\right|\tau=\frac{\mathrm i}{l}
\right)\right]\right|_{x=0}
\,\,\,\,\,\,j=0,1,\ldots,l-1;\,\,k=0,1,\,\ldots,
\end{equation}
These states are not orthogonal and, surprisingly, are 
discrete approximations of the continuum harmonic oscillator states. 

The ground state, ${\varphi}_{0}(j)$, is a Gaussian and the action of 
${\cal U}_{{\cal R}_0}$ on it is maximally dispersive. 
However, since ${\cal U}_{{\cal R}_0}$ has 
a finite period, the evolution of the ground state is periodic.

We end with some open problems. 
The above findings suggest that the naive continuum ({\em not} classical) limit of the eigenstates of ${\cal U}_{{\cal R}_0}$
doesn't lead to sensible results for integer sequences, $l_n=p^n$, for a 
{\em fixed} prime $p$ and $n=1,\ldots$. For most of the extended 
states, this limit exists in the Hilbert space ${\Bbb H}_{\Gamma}$ 
and is zero, since $\sum_{m=1}^{l} |c_m|^2=1$ and $c_m\approx O(1/\sqrt{l})$;
it may be possible to find suitable sequences, ${\cal U}_{{\cal R}_0^{r_n}}$, such that only some localized states survive in the limit. 
On the other hand, from the construction of the $p-$adic numbers, ${\Bbb Q}_p$ and $SL(2,{\Bbb Q}_p)$~\cite{st}, it is known that there does exist 
another ``continuum'' limit, in the $p-$adic numbers, which is called {\em projective} and is 
related to $p-$adic quantum mechanics for $p^n$, $n\to\infty$ (cf. Y. Meurice 
in ref.~\cite{st} and references therein). However the relation between the 
finite fields and the $p-$adic numbers is far from obvious and the relation
between our construction and that valid for the $p-$adics not known at the moment.  

For higher dimensional phase 
spaces the construction of the metaplectic 
representations of the symplectic group, $Sp(2D, l)$ (where 
$D$ is the dimension of the space), follows similar lines.

Regarding ``practical'' applications, 
the eigenstates of ${\cal U}_{{\cal R}_0}$ can be 
used to construct finite, orthogonal sets of wavelets over 
finite fields~\cite{afetal}, appropriate for analyzing {\em local} 
time-frequency or position-scale statistics of images. Another area 
is coding theory (especially cryptography). Some standard codes are 
linear or polynomial transformations over finite fields~\cite{sigproc2}. 
Our present work could be useful in ``quantizing'' linear codes or 
writing codes executable by quantum computers~\cite{feynman} as well as 
implementing algorithms for specifically quantum computation~\cite{shor}.

{\sl Acknowledgements:} The work of G. G. A. is partially supported by 
EEC~grants SCI-430C and CHRX-CT92-0063. E. G. F. and S. N. acknowledge 
the warm hospitality at the Laboratoire de Physique Th\'eorique of 
the Ecole Normale Sup\'erieure.


\begin{thebibliography}{99}
\bibitem{numbertheory} H. Weyl, ``Algebraic Theory of Numbers'' in
{\sl Annals of Mathematical Studies}, {\bf 1} (1940),Princeton Univ. Press.

G. H. Hardy and E. M. Wright, {\sl An Introduction to the Theory of Numbers}, 
5th edition, Oxford Science Public, Clarendon Press (1978). 

S. Lang, {\sl Algebraic Number Theory}, Addison-Wesley, N. Y. (1970). 

\bibitem{qc2}
V. I. Arnold, ``Mathematical Methods of Classical Mechanics'', Springer Graduate Texts in Mathematics, {\bf 60} Springer-Verlag (1978). 

V. I. Arnold and A. Avez, {\sl Ergodic Problems of Classical Mechanics}, 
Benjamin, N. Y. (1968). 
 
\bibitem{qc3}
F. Vivaldi, ``Geometry of Linear Maps over Finite Fields'', 
{\sl Nonlinearity} {\bf 5} (1992) 133. 
 
\bibitem{qc4} 
J. Hannay and M. V. Berry, {\sl Physica} {\bf D1} (1980) 267.

M. V. Berry, {\sl Proc. R. Soc. Lond.} {\bf A473} (1987) 183.  

\bibitem{qc5} 
N. L. Balazs and A. Voros, {\sl Phys. Rep.} {\bf 143C} (1986) 109.

B. V. Chirikov, F. M. Izrailev and D. Shepelyansky, 
{\sl Physica} {\bf D33} (1988) 77.
 
J. Ford, G. Mantica and G. H. Ristow, {\sl Physica } {\bf D50} (1991) 493. 

P. Leboeuf and A. Voros in {\sl Quantum Chaos}, eds. G. Casati and B. V. Chirikov, Cambridge Univ. Press, Cambridge (1993).

\bibitem{loc} J. B. Sokoloff, {\sl Phys. Rep.} {\bf 126C} (1985) 184.

\bibitem{rct1} A. Cappelli, C. Itzykson and J.-B. Zuber, {\sl Comm. Math. Phys.} {\bf 113} (1987) 1. 

\bibitem{rct2} A. Coste and T. Gannon, {\sl Phys. Lett.} {\bf B323} (1994) 316.

\bibitem{st} P. G. O. Freund and E. Witten {\sl Phys. Lett.} {\bf B199} (1987) 191; 
Ph. Ruelle, E. Thiran, D. Verstegen and J. Weyers, {\sl J. Math. Phys.} {\bf 30} (1989) 2854; 
Y. Meurice, {\sl Int. J. Mod. Phys. } {\bf A4} (1989) 5133; 
Y. Meurice, {\sl Comm. Math. Phys. } {\bf 135} (1991) 303; 
Y. Meurice, {\sl Phys. Lett. } {\bf B245} (1990) 99; 
L. Brekke and P. G. O. Freund, ``$p-$adic Numbers in Physics'', 
{\sl Phys. Rep.} {\bf 233C} (1993) 1.   

\bibitem{leshouches} Les Houches Conf., {\sl From Number Theory to Physics}, 
eds. M. Waldschmidt, P. Moussa, J.-M. Luck and C. Itzykson, Springer-Verlag 
(1992). 

\bibitem{sigproc1} M. R. S. Schroeder, {\sl Number Theory in Science and 
Communications}, corr. 2nd printing, Springer series in Information Sciences, 
Springer-Verlag (1989).

\bibitem{sigproc2} B. Lidl and H. Niederreiter, ``Finite Fields and their 
Applications'', {\sl Encyclopedia of Mathematics} {\bf 20}, 
Cambridge University Press, Cambridge, (1984). 

\bibitem{bh} 
J. Bekenstein, 
``Black Hole Thermodynamics'', {\sl Phys. Today} (1980).;
``Do we understand black hole entropy?'', 
gr-qc/9409015', talk at 7th Marcel Grossmann Symposium on General Relativity. 


G. 't Hooft, {\sl Nucl. Phys.} {\bf B335} (1990) 138; 
{\sl Nucl. Phys. } {\bf B342} (1990) 471.

L. Susskind, L. Thorlacius and J. Uglum, {\sl Phys. Rev.} {\bf D49} (1993) 3743. 

J. Polchinski and A. Strominger, ``A Possible Resolution of the Black Hole 
Information Paradox'', {\sl Phys. Rev.} {\bf D50} (1994) 7403.

Y. Kiem, H. Verlinde and E. Verlinde, ``Black Hole Horizons and 
Complementarity'', hep-th/9502074.

F. Larsen and F. Wilczek, ``Renormalization of Black Hole Entropy and the 
Gravitational Coupling Constant'', hep-th/9506006.

\bibitem{21-24} 
G. Veneziano, {\sl Europhys. Lett.} {\bf 2} (1986) 199.

D. J. Gross and P. Mende, {\sl Phys. Lett.} {\bf B197} (1987) 129; 
{\sl Nucl. Phys. } {\bf B303} (1988) 407. 

D. Amati, M. Ciafaloni and G. Veneziano, {\sl Phys. Lett. } {\bf B197} (1987) 81. 

K. Konishi, G. Paffuti and P. Provero, {\sl Phys. Lett. } {\bf B276} (1990) 276.

\bibitem{afetal} G. G. Athanasiu and E. G. Floratos, {\sl Nucl. Phys. } {\bf B425} (1994) 343.

\bibitem{bi} R. Balian and C. Itzykson, {\sl C. R. Acad. Sci. Paris} {\bf 303},
 s\'erie I(16), (1986) 773.

\bibitem{cartier} P. Cartier, ``Quantum Mechanical Commutation Relations and 
Theta Functions'' in {\sl Proc. Symp. Pure Mathematics}, {\bf 9}: 
{\sc Algebraic--Discontinuous Groups}, AMS, Providence, RI (1966). 

\bibitem{Weyl} 
H. Weyl, {\sl The Theory of Groups and Quantum Mechanics}, 
Dover, N. Y. (1931). 

J. Schwinger, {\sl Proc. Nat. Acad. Sci.} {\bf 46} (1960) 257; 544; 883. 

\bibitem{mumford} D. Mumford, {\sl Tata Lectures on Theta}, {\bf I--III}, 
Birkh\"auser, N. Y. (1986). 

\bibitem{weil} A. Weil, {\sl Acta Mathematica} {\bf 111} (1964) 143.

\bibitem{tanaka} S. Tanaka, {\sl Osaka J. Math. } {\bf 3} (1966) 229.
 
\bibitem{tanaka2} S. Tanaka, {\sl Osaka J. Math. }{\bf 4} (1967) 65.

\bibitem{evans} R. J. Evans, ``Hermite Character Sums'', {\sl Pacific J. Math.} 
{\bf 122} (1986) 357.

\bibitem{auslander} S. Auslander and P. Tolmieri, {\sl Bull. Am. Math. Soc.} 
{\bf 1} (1979) 847. 

B. C. Berndt and R. J. Evans, {\sl Bull. Am. Math. Soc.} {\bf 5} (1981) 107.

\bibitem{mehta} M. L. Mehta, {\sl Matrix Theory} , Editions  de Physique, 
Paris (1989). 

\bibitem{feynman}
R. P. Feynman, ``Quantum Mechanical Computers'', {\sl Foundations of Physics}, 
{\bf 16} (1986) 507.

D. Deutsch, ``Quantum Computational Networks'', {\sl Proc. R. Soc. Lond.} 
{\bf A425} (1989) 73.

S. Loyd, ``Envisioning a quantum supercomputer'', {\sl Science}, {\bf 263}
(1994) 695.

\bibitem{shor}
P. W. Shor, ``Algorithms for Quantum Computation: Discrete Logarithms and 
Factoring'' in {\sl Proc. IEEE Computer Society Press}, Nov. 1994, p.124.




\end{thebibliography}
\end{document}